\documentclass[twoside]{article}
\usepackage{fleqn,espcrc2}
\font\sqi=cmssq8
\def\DR{\rm I\kern-1.45pt\rm R}
\def\DC{\kern2pt {\hbox{\sqi I}}\kern-4.2pt\rm C}
\def\DH{\rm I\kern-1.5pt\rm H\kern-1.5pt\rm I}

\title{K\"ahler geometry and  SUSY  mechanics.}
\author{Stefano Bellucci\address{INFN, Laboratori Nazionali di Frascati,
P.O. Box 13, I-00044, Frascati, Italy} and Armen
Nersessian$\;^{\rm a}\;$\address{JINR, 
Bogolyubov Laboratory of Theoretical Physics,
 Dubna, 141980  Russia}
\address{Yerevan State University, A.Manoogian, 1, Yerevan,
375025 Armenia}}
\begin{document}
\begin{abstract}
\noindent
We present two examples of SUSY mechanics related with K\"ahler geometry. 
The first system is the  $N=4$ supersymmetric one-dimensional
sigma-model  proposed in {\tt hep-th/0101065}.
Another system is the $N=2$ SUSY mechanics whose phase space is the 
external algebra of an arbitrary K\"ahler manifold.
The relation of these models 
 with  antisymplectic geometry is discussed.
\end{abstract}
\maketitle
\setcounter{section}{0}
\section{Introduction}
Supersymmetric mechanics attracts permanent interest
since its introduction \cite{witten}. However,
studies focussed mainly on the $N=2$ case, and the
most important case of $N=4$ mechanics did not receive enough
attention,
though some interesting observations were made about this subject:
let us mention that the most general
 $N=4, D=1,3$  supersymmetric mechanics described by real superfield
actions  were  studied in Refs. \cite{ikp,is} respectively, and those
in arbitrary $D$ in Ref. \cite{dpt};
in \cite{bp} $N=4, D=2$ supersymmetric mechanics described
by chiral superfield actions were considered;
the general study of supersymmetric mechanics with arbitrary
$N$ was performed recently in Ref. \cite{hull}.
In the Hamiltonian language classical supersymmetric mechanics
can be formulated
in terms of  superspace equipped with some
supersymplectic structure (and  corresponding non-degenerate
Poisson brackets). After quantization
the odd coordinates are replaced by the generators
of Clifford algebra. It is easy to verify that the
minimal dimension of phase superspace, which allows to describe
a $D-$dimensional supersymmetric mechanics
with {\sl nonzero} potential terms, is
$(2D|2D)$, while supersymmetry specifies
both the admissible sets of configuration spaces and potentials.
In our recent paper \cite{bn} we proposed the $N=4$
 supersymmetric one-dimensional
sigma-models (with and without central charge) on  K\"ahler manifold
with $(2d|2d)_{\DC}$-dimensional  phase space.
We have shown that the constructed mechanics
can be obtained by dimensional
reduction from  $N=2$ supersymmetric $(1+1)-$dimensional sigma-models
 by Alvarez-Gaum\'e and Freedman \cite{agf};
in the simplest case of $d=1$  and in the absence of central
charge these systems coincide with the $N=4$ supersymmetric mechanics
described by the chiral superfield action \cite{bp}.

In  Section 2 we present the $N=4$ one-dimensional
 supersymmetric sigma-model constructed in Ref. \cite{bn}. 

In  Section 3  we present  a new  model of $N=2$ supersymmetric
 mechanics with  phase space corresponding the  external algebra of 
a K\"ahler manifold. This construction  seems to be the most general
 $N=2$ SUSY mechanics  $(2D|2D)-$dimensional  phase space.
 We also consider the relation of presented system
with the antisymplectic
geometry,  in the context of the old problem  suggested 
by D.V.Volkov et al. \cite{vpst}.

\section{Sigma-models with   $N=4$ SUSY.}

In order to get  a one-dimensional
$N=4$ supersymmetric sigma-model   with  $(2D|2D)-$dimensional
 phase superspace
 one should require that
the target space $M_0$ is a K\"ahler manifold
$(M_0, g_{a\bar b}dz^ad{\bar z}^{\bar b})$,
$g_{a\bar b}=\partial^2 K(z,\bar z)/\partial z^a\partial{\bar z}^b$.
This restriction follows also from the
considerations of superfield actions:
indeed, the $N-$extended supersymmetric
mechanics obtained from the
action depending on $D$  real superfields,
have a $(2D|ND)_{\DR}$-dimensional symplectic manifold,
whereas those obtained from the action depending on  $d$
chiral superfields have a $(2d|Nd/2)_{\DC}-$dimensional
phase space, with the configuration space being a
$2d-$dimensional K\"ahler manifold.

In that case the phase superspace
can be equipped  by the supersymplectic structure
\begin{equation}
\begin{array}{c}
\Omega=\omega_0-i\partial{\bar\partial}{\bf g}=\\
=d\pi_a\wedge dz^a+ d{\bar\pi}_a\wedge d{\bar z}^a+\\
+R_{a{\bar b}c\bar d}\eta^a_i\bar\eta^b_i dz^a\wedge d{\bar z}^b+
g_{a\bar b}D\eta^a_i\wedge{D{\bar\eta}^b_i}
\end{array}
\label{ss}\end{equation}
where
\begin{equation}
\begin{array}{c}
{\bf g}=ig_{a\bar b}\eta^a\sigma_0{\bar\eta}^b,\\
D\eta^a_i=d\eta^a_i+\Gamma^a_{bc}\eta^a_i dz^a,\quad i=1,2
\end{array}\end{equation}
while $\Gamma^a_{bc},\; R_{a\bar b c\bar d}$ are
 respectively the connection and curvature of the K\"ahler structure,
 the odd coordinates $\eta^a_i$ belong to the external algebra
$\Lambda(M_0)$, i. e. transforms as  $dz^a$.
This symplectic structure becomes canonical
in the coordinates $(p_a,\chi^k)$
\begin{equation}
\begin{array}{c}
p_a=\pi_a-\frac{i}{2} \partial_a{\bf g},
\quad\chi^m_i={\rm e}^m_b\eta^b_i:\\
\Omega=dp_a\wedge d z^a +d{\bar p}_{\bar a}\wedge d{\bar z}^{\bar a}
+d\chi^m_i\wedge d{\bar\chi}^{\bar m}_i,
\end{array}
\label{canonical}\end{equation}
where ${\rm e}^m_a$ are the einbeins of the  K\"ahler structure:
${\rm e}^m_a\delta_{m\bar m}{\bar{\rm e}}^{\bar m}_{\bar b}=g_{a\bar b}.$
So to quantize this model, one chooses
$$ 
{\hat p}_a=-i\frac{\partial}{\partial z^a},\;
 {\hat{\bar  p}}_{\bar a}=-i\frac{\partial}{\partial \bar z^{\bar a}},\;
[{\hat\chi}^m_i,{\hat{\bar\chi}}^{\bar n}_j]_+
=\delta^{m\bar n}\delta_{ij}.
$$
The corresponding Poisson brackets are defined
by the following non-zero
relations (and their  complex-conjugates)
$$
\begin{array}{c}
\{\pi_a, z^b\}=\delta^b_a,\quad
\{\pi_a,\eta^b_i\}=-\Gamma^b_{ac}\eta^c_i,\\
\{\pi_a,\bar\pi_b\}=-R_{a\bar b c\bar d}\eta^c_k{\bar\eta}^d_k,\quad
\{\eta^a_i, \bar\eta^b_j\}=g^{a\bar b}\delta_{ij}.
\end{array}
$$

To construct on this phase superspace
the Hamiltonian mechanics  with  standard $N=4$ supersymmetry algebra
 \begin{equation}
\begin{array}{c}
\{Q^+_i,Q^-_j\}=\delta_{ij}{\cal H},\\
\{Q^\pm_i,Q^\pm_j\}=\{Q^\pm_i, {\cal H}\}=0,\quad i=1, 2,
\end{array}
\label{4sualg}\end{equation}
let us choose the  supercharges  given by the functions
 \begin{equation}
Q^+_1=\pi_a\eta^a_1+ iU_{\bar a}{\bar \eta}^{\bar a}_2,\quad
Q^+_2=\pi_a\eta^a_2- iU_{\bar a}{\bar \eta}^{\bar a}_1.
\label{4SUSY}\end{equation}
Then, calculating the commutators (Poisson brackets) of these
functions, we get that the supercharges (\ref{4SUSY})  belong
to the superalgebra (\ref{4sualg})
when the functions $U_a, {\bar U}_{\bar a}$
are  of the form
\begin{equation}
U_a(z)=\frac{\partial U(z)}{\partial z^a},\quad
{\bar U}_{\bar a}(\bar z )=
\frac{\partial {\bar U}({\bar z})}{\partial {\bar z}^a},
\end{equation}
while   the Hamiltonian reads
\begin{equation}
\begin{array}{c}
{\cal H}=g^{a{\bar b}}(\pi_a{\bar\pi}_b+
{U}_a{\bar U}_{\bar b})-\\
-iU_{a;b}\eta^a_1\eta^{b}_2
+i{\bar U}_{\bar a;\bar b}{\bar\eta}^{\bar a}_1{\bar\eta}^{\bar b}_2
-R_{a\bar b c\bar d}\eta^a_1\bar\eta^b_1\eta^a_2\bar\eta^d_2,
\end{array}
\label{4SUHam}\end{equation}
 where
$ U_{a;b}\equiv \partial_a\partial_b U-\Gamma^c_{ab}\partial_cU$.

The constant of motion  counting the number of fermions, reads
\begin{equation}
{\cal F}=ig_{a\bar b}\eta^a\sigma_3{\bar\eta}^{\bar b}:\;\;
 \{Q^\pm_i, {\cal F}\}=\pm i Q^\pm_i,
\label{f}\end{equation}

Notice that the above-presented $N=4$ SUSY
mechanics for the simplest case, i.e. $d=1$,
was obtained by Berezovoy and Pashnev \cite{bp}
from the chiral superfield action
 \begin{equation}
 {\cal S}=\frac 12\int K(\Phi,\bar\Phi)
+2\int U(\Phi)+2\int{\bar U}(\bar\Phi)
\end{equation}
where $\Phi$ is  a chiral superfield.
It seems to be obvious that a similar action depending on $d$
chiral superfields will generate the above-presented $N=4$ SUSY
mechanics.\\

Let us consider a generalization of the above system,
which possesses $N=4$  supersymmetry with
central charge
\begin{equation}
\begin{array}{c}
\{\Theta^+_i,\Theta^-_j\}=
\delta_{ij}{\cal H}+{\cal Z}\sigma^3_{i{j}},\quad
\{\Theta^\pm_i,\Theta^\pm_j\}=0,
\\
\{{\cal Z}, {\cal H}\}= \{{\cal Z},\Theta^\pm_k\}=0.
\end{array}
\label{csa}\end{equation}
For this purpose one introduces
the supercharges
\begin{equation}
\begin{array}{c}
 \Theta^+_1=\left(\pi_a+iG_{,a}(z,\bar z)\right)\eta^a_1 +
i {\bar U}_{,\bar a}({\bar z}){\bar \eta}^{\bar a}_2,\\
 \Theta^+_2=\left(\pi_a-iG_{,a}(z,\bar z)\right)\eta^a_2 -
i {\bar U}_{,{\bar a}}({\bar z}){\bar \eta}^{\bar a}_1,
\end{array}
\end{equation}
where the real function $G(z,\bar z)$  obeys the conditions
\begin{equation}
\partial_a\partial_b G
+\Gamma^c_{ab}\partial_c G=0,\;
{ G}_{,a}(z,\bar z)g^{a\bar b}{\partial}_{\bar b}{\bar U}({\bar z})=0.
\label{killing}\end{equation}
The first equation in (\ref{killing}) is nothing
but the Killing equation
of  the underlying K\"ahler structure
(let us remind, that the isometries of the K\"ahler structure are
Hamiltonian holomorphic vector fields) given by the vector
\begin{equation}
{\bf G}=G^a(z)\partial_a+{\bar G}^a({\bar z}){\bar\partial}_a,\quad
G^a=ig^{a\bar b}{\bar \partial}_b G.
\end{equation}
The second equation means that the vector field ${\bf G}$
leaves the holomorphic function invariant
$${\cal L}_{\bf G}U=0\;\Rightarrow \;G^a(z)U_a(z)=0.$$

Calculating the Poisson brackets of these supercharges,
we get explicit
expressions for the Hamiltonian
\begin{equation}
\begin{array}{c}
{\cal H}\equiv
g^{a{\bar b}}\left(\pi_a{\bar\pi}_{\bar b}+ G_{,a}G_{{\bar b}}
+{ U}_{,a}{\bar U}_{,\bar b}\right)- \\
-iU_{a;b}\eta^a_1\eta^{b}_2 +
i{\bar U}_{\bar a;\bar b}{\bar\eta}^{\bar a}_1{\bar\eta}^{\bar b}_2
+\frac 12 G_{a\bar b}(\eta^a_k\bar\eta^{\bar b}_k) -\\
-R_{a\bar b c\bar d}\eta^a_1\bar\eta^b_1\eta^c_2\bar\eta^d_2 
\end{array}
\end{equation}
and the central charge
\begin{equation}
\begin{array}{c}
{{\cal Z}}=i(G^a\pi_a+G^{\bar a}{\bar\pi}_{\bar a})+\frac{i}{2}
\partial_a{\bar\partial}_{\bar b}G(\eta^a{\sigma_3}\bar\eta^{\bar b}).
\end{array}
\label{z}\end{equation}
It can be checked by a straightforward calculation that
the function ${\cal Z}$ indeed belongs to the center of
the superalgebra (\ref{csa}).
The scalar part of each  phase
with  standard $N=2$ supersymmetry can be interpreted
as a particle  moving on the K\"ahler  manifold
in the presence of an external magnetic field with strength
$F=iG_{a\bar b}dz^a\wedge d{\bar z}^{\bar b}$ and in the
potential field $U_{,a}(z)g^{a\bar b}{\bar U}_{,\bar b}(\bar z)$.

Assuming that $(M_0, g_{a\bar b}dz^a d{\bar z}^b)$ is the
hyper-K\"ahler metric and that $U(z)+{\bar U}({\bar z})$
is a tri-holomorphic function while the function
$G(z,\bar z)$ defines a tri-holomorphic Killing vector,
one should get the $N=8$ supersymmetric one-dimensional sigma-model.
In that case instead of the
phase with standard $N=2$ SUSY arising in the
K\"ahler case, we shall get the phase with standard $N=4$ SUSY.
The latter system can be viewed as a particular
case of $N=4$ SUSY mechanics describing the
low-energy dynamics of monopoles and dyons in
$N=2,4$ super-Yang-Mills theory \cite{gk}.
Notice that, in contrast to the $N=4$ mechanics suggested
in the mentioned
papers, in the above-proposed (hypothetic) construction
also the four hidden supersymmetries could be explicitly written.

The Lagrangian of the system is of the form
\begin{equation}
\begin{array}{c}
{\cal L}=
g_{a{\bar b}}\left({\dot z}^a{\dot{\bar z}}^{b}+
\frac 12 \eta^a_k\frac{D{\bar\eta}^{\bar b}_k}{d\tau}+
\frac 12 \frac{D\eta^{a}_k}{d\tau}\bar\eta^{\bar b}\right)-\\
-g^{a{\bar b}}(G_aG_{\bar b}+{U}_a{\bar U}_{\bar b})+\\
+iU_{a;b}\eta^a_1\eta^{b}_2-
i{\bar U}_{\bar a;\bar b}{\bar\eta}^{\bar a}_1{\bar\eta}^{\bar b}_2+
R_{a\bar b c\bar d}\eta^a_1\bar\eta^b_1\eta^a_2\bar\eta^d_2.
\end{array}
\end{equation}
So, we get the Lagrangian  for a one-dimensional sigma-model
 with four exact real supersymmetries. It can be straightforwardly obtained
 by the dimensional reduction of the $N=2$ supersymmetric
$(1+1)$ dimensional sigma-model by
 Alvarez-Gaum\'e and Freedman \cite{agf}
(the mechanical counterpart of this system without
potential term was constructed in \cite{macfarlane}).

\section{$N=2$ SUSY mechanics with K\"ahler phase space}
Let us  consider a supersymmetric mechanics
 whose phase superspace is the external
algebra of the K\"ahler manifold $\Lambda({M})$, where
$\left({M},\; g_{A\bar B}(z,\bar z)dz^A d{\bar z}^{\bar B}\right)$
 plays the role of the phase space
of underlying Hamiltonian mechanics.
The phase superspace is
$(D|D)_{\DC}-$ dimensional  supermanifold
 equipped by  the K\"ahler
 structure \cite{knkahler}
\begin{equation}
\begin{array}{c}
\Omega=i\partial \bar\partial
\left(K(z,\bar z)-ig_{A\bar B}\theta^A{\bar\theta}^{\bar B}
\right)=\\
=i(g_{A\bar B}+i
R_{A\bar BC\bar D}\theta^C\bar\theta^{\bar D})dz^A\wedge
 d{\bar z}^{\bar B}
\\+g_{A\bar B}D\theta^A\wedge D{\bar\theta}^{\bar B},
\end{array}
\label{ssg}\end{equation}
where
$D\theta^A=d\theta^A +
\Gamma^A_{BC}\theta^B dz^C$, and
 $\Gamma^A_{BC}$, $R_{A\bar BC\bar D}$ are respectively
the Cristoffel symbols and curvature tensor of the underlying
K\"ahler metrics $g_{A\bar B}=\partial_A\partial_{\bar B}K(z,\bar z)$.

The corresponding Poisson bracket can be presented in the form
\begin{equation}
\begin{array}{c}
\{\quad,\quad\}=i{\tilde g}^{A\bar B}\nabla_A
\wedge{\bar\nabla}_{\bar B}+g^{A\bar B}\frac{\partial}{\partial \theta^A}
\wedge\frac{\partial}{\partial{\bar\theta}^{\bar B}}
\end{array}
\end{equation}
where
$$
\nabla_A=\frac{\partial}{\partial z^A}-
\Gamma^C_{AB}\theta^B\frac{\partial}{\partial\theta^C},
$$
and
$$
{\tilde g}^{-1}_{A\bar B}=(g_{A\bar B}+
 iR_{A\bar BC\bar D}\theta^C\bar\theta^{\bar D}).
$$

On this phase superspace one can immediately construct
a mechanics  with standard $N=2$ supersymmetry
 \begin{equation}
\{Q_+,Q_-\}={{\cal H}},\quad\{Q_\pm,Q_\pm\}=\{Q_\pm, {\cal H}\}=0,
\label{2psualg}\end{equation}
 given by the supercharges
\begin{equation}
\begin{array}{c}
Q^0_+=\partial_A K(z, \bar z)\theta^A,\quad
Q^0_-=\partial_{\bar A}K(z, \bar z){\bar\theta}^{\bar A}
\end{array}
\label{q0}\end{equation}
where $K(z,\bar z)$
is  the K\"ahler potential 
 of the underlying K\"ahler structure, defined up to holomorphic and 
anti-holomorphic functions,
$$K(z, \bar z)\to K(z,\bar z)+f(z)+{\bar f}(\bar z).$$
The Hamiltonian of the system reads
\begin{equation}
\begin{array}{c}
{{\cal H}}_0=g^{A\bar B}\partial_A K \partial_{\bar B}K
-ig_{A\bar B}\theta^A{\bar\theta}^{\bar B}+\\
+i\theta^CK_{C;A}{\tilde g}^{A\bar B}K_{\bar B;\bar D}{\bar\theta}^{\bar D}
\end{array}
\label{h0}\end{equation}
where
$K_{A;B}=\partial_A\partial_BK-\Gamma^C_{AB}\partial_CK$.\\

One also consider another mechanics with standard $N=2$ SUSY  whose
 supercharges  are given by the expressions
\begin{equation}
\begin{array}{c}
Q^c_+=\partial_A G(z, \bar z)\theta^A,\quad
Q^c_-=\partial_{\bar A}G(z, \bar z){\bar\theta}^{\bar A}
\end{array}
\label{c0}\end{equation}
where $G(z,\bar z)$
is  the Killing potential of the underlying K\"ahler structure,
$$
\begin{array}{c}
 {\partial_A\partial_B}G-\Gamma^C_{AB}\partial_C G=0,\\
G^A(z)=g^{A\bar B}\partial_{\bar B}G(z,\bar z).
\end{array}
$$
In this case the  Hamiltonian of system reads
\begin{equation}
\begin{array}{c}
{{\cal H}^c}=g_{A\bar B} G^A G^{\bar B}+
 i{\bar\theta}^{\bar D}G_{A{\bar D}}{\tilde g}^{A\bar B}
G_{C\bar B}\theta^C,
\end{array}
\label{hc}\end{equation}
where $G_{A\bar B}=\partial_A{\partial}_{B}G(z,\bar z)$.

The commutators of the supercharges in  these particular examples
read:
\begin{equation}
  \{Q^c_\pm, Q^0_\pm\}={\cal R}_\pm,\quad
\{Q^c_\pm, Q^0_\mp\}={\cal Z},
\end{equation}
where
\begin{equation}
\begin{array}{c}
 {\tilde{\cal Z}}\equiv G(z,\bar z)+
iG_{A\bar B}(z,\bar z)\theta^A{\bar\theta}^{\bar B},\\
{\cal R}_+=i\theta^CK_{C;A}{\tilde g}^{A\bar B}G_{\bar B; D}
{\theta}^{ D},\\
{\cal R}_-=\bar{\cal R}_+\;\;. 
\end{array}
\label{Z}\end{equation}
Hence, introducing the superschages 
\begin{equation}
\Theta_\pm=Q^0_\pm \pm iQ^c_\mp, 
\end{equation}
  we can define  $N=2$ SUSY mechanics specified by 
the presence of central charge ${\cal Z}$:
\begin{equation}
\begin{array}{c}
\{\Theta_+,\Theta_-\}={\tilde{\cal H}},\quad
 \{\Theta_\pm, \Theta_\pm\}=\pm i{\cal Z}\\
\{{\cal Z},\Theta_\pm\}=0,\;\;-\{{\tilde{\cal H}},\Theta_\mp\}=0,\;\;
\{{\cal Z}, {\tilde{\cal H}}\}=0.
\end{array}
\end{equation}
The Hamiltonian of this 
generalized mechanics is defined by the expression
\begin{equation}
{\tilde{\cal H}}={\cal H}_0+{\cal H}_c +i{\cal R}_+ -i{\cal R}_-.
\end{equation}
A ``fermionic number" is of the  form:
\begin{equation}
{\tilde {\cal F}}=ig_{AB}\theta^A{\bar\theta}^{\bar B}:
\;
 \{ {\tilde{\cal F}},\Theta_\pm,\}=\pm i \Theta_\pm \; .
\end{equation}
Choosing the K\"ahler manifold to be a special type of the 
(co)tangent bundle of 
some K\"ahler manifold, one can provide the system 
by the additional pair of supercharges, recovering the above-constructed
 model of sigma-model with standard $N=4$ SUSY.

The phase space of the system under consideration can be equipped,
in addition to the Poisson bracket corresponding to (\ref{ssg}),
with the  antibracket (odd Poisson bracket)
 associated with the odd K\"ahler potential
$K_\alpha=\alpha Q^0_+ +{\bar\alpha}Q^0_-$, $ \alpha=1, i$:
\begin{equation}
\{\quad,\quad\}^{(\alpha)}_1=
\alpha{g^{{\bar A} B}}\nabla_{\bar A}\wedge
\frac{\partial}{\partial\theta^B} +c.c.\quad.
\label{antib}\end{equation}
It is easy to observe, that the following equality holds \cite{knkahler}:
\begin{equation}
\{{\tilde {\cal Z}}, \;\;\}=\{Q^\alpha, \;\;\}^{(\alpha)}_1,
\label{volk}\end{equation}
where
\begin{equation}
\begin{array}{c}
Q^\alpha=\alpha Q^c_+  +{\bar\alpha}Q^c_- .
\end{array}
\end{equation}
Hence, the anti-symplectic structure generated by
supercharges  of the $N=2$ SUSY mechanics(\ref{q0}), (\ref{h0}) 
define the pair of 
anti-Hamiltonian structures for the Hamiltonian field corresponding to central 
charge (\ref{Z}), while
 the supercharges of the SUSY mechanics (\ref{c0}),(\ref{hc})
play the role of corresponding Hamiltonians.

The following relation holds:
\begin{equation}
\{G+{\cal F},G+{\cal F}\}^{(\alpha)}_1=2Q^\alpha,
\label{lie}\end{equation}
which can be interpreted as  
$N=1$ supersymmetry of the 
anti-Hamiltonian mechanics given by  the (odd) Hamiltonian $Q$.
However, this supersymmetry is {\it trivial} (or false): 
it  does not changes  the initial Hamilonian dynamics,
 generated on the base manifold by the Killing potential $G$.
On the other hand,  squaring the   odd Hamiltonian $Q$ under 
the even Poisson bracket 
(corresponding  to  squaring the corresponding  quantum-mechanical 
supercharge  ${\hat Q}$)  yields  the Hamiltonian (\ref{hc}).
Hence, in the   $N=2$ SUSY mechanics given by the Hamiltonian
(\ref{hc}), allows us to get the ``square root'' 
by the use of the antibracket.

Notice that the supermanifolds provided by the even and odd
 symplectic (and K\"ahler) structures, and particularly Eq. \ref{volk},
 were studied in cinnection with the problem of
 describing  the   supersymmetric mechanics
in terms of antibrackets \cite{knkahler,knj},
which was  considered for the first time by D.V.Volkov et al. 
 \cite{vpst}.
Let us remind that until the 1980's the odd Poisson brackets (antibrackets)
had no any applications in   theoretical physics, due to the nontrivial 
Grassmann grading, and  the absence of consistent quantization 
schemes. 
This situation drastically changed in  1981, when
 I.A.Batalin and G.A.Vilkovisky suggested a covariant 
Lagrangian BRST quantization scheme
(which is presently  known as BV-formalism) \cite{bv}, 
whose key ingredient was the odd Poisson bracket.
A bit later, in 1983, D.V.Volkov claimed, that
the antibrackets can be considered, due to their non-zero Grassmann grading,
as the ``square root'' of usual super-Poisson brackets relating the bosonic 
and fermionic components of super-spinors \cite{volkov0}.
The study of the possibility of an antibracket formulation of
 supersymmetric Hamiltonian systems, i.e.  Eq. \ref{volk},
was one of the steps of that program.   
 Later,  the supersymmetric mechanics, which are Hamiltonian 
with respect to both even and odd Poisson brackets,
were found to be useful in  equivariant cohomology,
 e.g.  for the  construction of equivariant
characteristic classes and the derivation of
localization formulae \cite{ec}. Indeed,  the vector field
(\ref{volk}) can be identified with the Lie derivative,
 if choose $\alpha=i$:
\begin{equation}
\{Q,\;\}^{(i)}_1=iG^A(z)\frac{\partial}{\partial z^A}+
iG^A_{,C}(z)\theta^C\frac{\partial}{\partial z^A}+ \; c.c.
\end{equation}
 The vector field
$\{{\tilde{\cal F}},\;\;\}^{(i)}_1$, 
corresponds to the external differential
\begin{equation}
\{{\cal F},\;\}^{(i)}_1=\theta^A\frac{\partial}{\partial z^A}+ \; c.c.,
\end{equation}
while 
$\{G,\;\;\}^{(i)}_1$
corresponds to the
 operator of inner  product
\begin{equation}
\{G,\;\}^{(i)}_1=iG^A(z)\frac{\partial}{\partial \theta^A}+\; c.c.\;\;.
\end{equation}
Hence, equipping the external algebra of the K\"ahler manifold
by  a  pair of antibrackets (corresponding to the choice
$\alpha=0,\pi/2$) one can 
describe the external calculus in terms of the pair of antibrackets.
One can observe that
$\{G+{\cal F},\;\;\}_1$ defines equivariant differential, while 
the function $G-{\cal F}$ defines the  equivariant Chern class;
the relation (\ref{lie})
corresponds to the well-known Lie identity
$di_X+i_Xd=L_X$, and so on.
The Lie derivative of odd  symplectic structure  along
the vector field $\{G+{\cal F},\;\;\}_1$ yields the equivariant
 even pre-symplectic structure, generating equivariant Euler classes of
the underlying K\"ahler manifold \cite{ec}.

Let us mention also another similarity between
 the system under consideration and the Lagrangian BRST quantization schemes.
The Batalin-Vilkovisky formalism  admits the 
 BRST-antiBRST-invariant extension
which is known, in its most general form, under the name of 
``triplectic formalism'' \cite{blt}.
This extension is formulated by the use of a pair of antibrackets,
 $\{\;,\;\}^\alpha_1$,  the pair of  nilpotent odd vector fields obeying
the compatibility conditions
\begin{equation}
\begin{array}{c}
(-1)^{(p(f)+1)(p(h)+1)}\{f,g\}^{\{\alpha}_1,h\}^{\beta\}}_1+\\
 +{\rm cicl. perm.}\;f,g,h=0\\
V^{\{\alpha}\{f,g\}^{\beta\}}_1=\\
=\{V^{\{\alpha}f,g\}^{\beta\}}_1+(-1)^{p(f)+1}\{f,
V^{\{\alpha}g\}^{\beta\}}_1\\
V^{\{\alpha}V^{\beta\}}=0.
\end{array}
\label{ta}\end{equation}
It is easy to see, that the antibrackets (\ref{antib}) 
 and the vector fields $V=\{{\cal F},\;\;\}_1$ corresponding to the choice
$\alpha=1, i$, form  a ``classical triplectic algebra" (\ref{ta}).
However, there is a crucial difference between the triplectic algebra 
arising in the context of BRST quantization and the above-presented one:
the antibrackets appearing in the triplectic formalism, are degenerate, 
while the above-presented ones are {\it nondegenerate}. On the other hand, 
``classical triplectic algebra'' corresponding to degenerate antibrackets,
is also related with K\"ahler geometry \cite{grig}. 

\section{Acknowledgments}The  authors are grateful to
S. Krivonos for  numerous enlightening discussions, to A. Galajinsky
for  useful comments
 and to the Organizers  of D.V.Volkov Memorial 
Conference ''Supersymmetry and Quantum 
Field Theory'' for inviting this contribution.
A.N. thanks   INFN for financial support and hospitality.


\begin{thebibliography}{99}
\bibitem{witten}E.~Witten, Nucl.~Phys. {\bf B188} (1981), 513;
{\it ibid.} {\bf B202} (1982), 253
\bibitem{ikp}E.~A.~Ivanov, S.~O.~Krivonos, A.~I.~Pashnev,
Class. Quant. Grav. {\bf 8}  (1991), 19
\bibitem{is}E.~A.~Ivanov, A.~V.~Smilga, Phys.~Lett. {\bf B257} (1991), 79

 V.~P.~Berezovoy, A.~I.~Pashnev, Class.~Quant.~Grav. {\bf 8} (1991), 2141
\bibitem{dpt}E.~E.~Donets et al.,
 Phys.~Rev. {\bf D61} (2000), 043512;
Phys.Lett. {\bf B484} (2000), 337

\bibitem{bp}V.~Berezovoy, A.~Pashnev, Class.~Quant.~Grav. {\bf 13} (1996),
 1699 
\bibitem{hull}C~.M.~Hull, {\sl The   Geometry of Supersymmetric
Quantum Mechanics}, {\tt hep-th/9910028}
\bibitem{bn}S.Bellucci, A. Nersessian, {\tt hep-th/0101065}
\bibitem{agf}L.~Alvarez-Gaume, D.~Freedman,
Comm. Math. Phys. {\bf 91}  (1983), 87;
for the quantum correction of sigma models to strings see e. g.\cite{io}
 and references therein
\bibitem{io}S. Bellucci, Z. Phys. C36 (1987), 229;
{\it ibid.} Prog. Theor. Phys. 79 (1988), 1288;
{\it ibid.} Z. Phys. C41 (1989), 631;
{\it ibid.} Phys. Lett. B227 (1989), 61; 
{\it ibid.} Mod. Phys. Lett. A5 (1990), 2253.
\bibitem{vpst}D.~V.~Volkov, A.~I.~Pashnev, V.~A.~Soroka, V.~I.~Tkach,
 JETP Lett. {\bf 44} (1986), 70
\bibitem{macfarlane}H.~W.~Braden, A.~J.~Macfarlane,
J.~Phys. {\bf A18} (1985), 2955
\bibitem{gk} D.~Bak, K.~Lee, P.~Yi, Phys.~Rev. {\bf D61} (2000), 045003;
{\it ibid.} {\bf D62} (2000), 025009

D.~Bak et al,  Phys.~Rev. {\bf D61} (2000), 025001;

J.~Gauntlett et al., Phys.~Rev. {\bf D61} (2000), 125012;
 {\tt hep-th/0008031}
\bibitem{knkahler}A.~P.~Nersessian, Theor.~Math.~Phys. {\bf 96} (1993), 866
\bibitem{knj}
O.~M.~Khudaverdian, A.~P.~Nersessian, J.~Math.~Phys. {\bf 32} (1991), 1938;
{\it ibid.} {\bf 34} (1993), 5533
\bibitem{ec} A.~P.~Nersessian,
JETP Lett. {\bf 58} (1993), 66;
Lecture Notes in Physics {\bf 524}, 90 ({\tt hep-th/9811110});
\bibitem{bv}I.~A.~Batalin, G.~A.~Vilkovisky,
 Phys.~Lett. {\bf B102} (1981), 27; Phys.~Rev. {\bf D28} (1983), 2567

\bibitem{volkov0}D.V.Volkov, JETP Lett., {\bf 38} (1983),615
\bibitem{blt}I.~A.~Batalin, P.~M.~Lavrov, I.~V.~Tyutin,
J.~Math.~Phys. {\bf 31} (1990), 1487;
{\it ibid.} {\bf 32} (1990), 532

I.~A.~Batalin, R.~Marnelius, A.~Semikhatov, Nucl.~Phys.
{\bf B446} (1995), 249

A.~Nersessian, P.~H.~Damgaard, Phys. Lett. {\bf B355} (1995), 150

\bibitem{grig}M. Grigoriev, A. M. Semikhatov, Theor. Math.  Phys.
{\bf 124} (2000), 1157 
\end{thebibliography}
\end{document}